\documentclass[structabstract]{aa}  
\usepackage{color,graphicx}
\usepackage{amsmath}
\usepackage{natbib}
\usepackage{txfonts}
\newcommand{\ed}[1]{#1}
\begin{document}
   \title{A unified view of coronal loop contraction and oscillation in flares}

   \author{A.~J.~B.~Russell\inst{1} \and P.~J.~A.~Sim\~oes\inst{2} \and L.~Fletcher\inst{2}}

   \institute{
   	      Division of Mathematics, University of Dundee, Nethergate, Dundee, DD1 4HN, Scotland, U.K.
             \and
             SUPA School of Physics and Astronomy, University of Glasgow, University Avenue, Glasgow, G12 8QQ, Scotland, U.K.\\
              \email{arussell@maths.dundee.ac.uk, paulo.simoes@glasgow.ac.uk, lyndsay.fletcher@glasgow.ac.uk}
    }

   \authorrunning{Russell et al.}
   \titlerunning{Contraction and oscillation of coronal loops}

   \date{Received; accepted}

 
  \abstract
   {Transverse loop oscillations and loop contractions are 
   commonly associated with solar flares,
   but the two types of  motion have traditionally been 
   regarded as separate phenomena.}
   {We present an observation of coronal loops contracting and oscillating following 
     onset of a flare.  We aim to explain why both behaviours are seen together
     and why only some of the loops oscillate.
    }
   {A time sequence of SDO/AIA 171 \r{A} images is analysed to identify 
   \ed{positions of coronal loops} following 
   the onset of M6.4 flare SOL2012-03-09T03:53.
   \ed{We focus on five loops in particular, all of which contract during the flare, with three of them oscillating as well.}
   A simple model is then developed for contraction and oscillation of a coronal loop.
    }
   {We propose that coronal loop contractions and oscillations can occur 
   in a single response to removal of magnetic energy from the corona.
   Our model reproduces the various types of loop motion observed
   and explains why the highest loops oscillate during their contraction while
   no oscillation is detected for the shortest contracting loops.
   The proposed framework suggests that loop motions
   can be used as a diagnostic for the removal of coronal magnetic energy by flares,
   while rapid decrease of coronal magnetic energy is a newly-identified 
   excitation mechanism for transverse loop oscillations.
    }
   {}

   \maketitle


\section{Introduction}\label{sec:intro}
This paper explores the relationship between contraction and oscillation 
of coronal loops in connection with solar flares.
It is motivated by the M6.4 flare SOL2012-03-09T03:53 and the associated coronal collapse,
which have been described in detail by \citet{Paper1}, hereafter referred to as Paper~I. 
Paper~I also noted the occurrence of transverse loop oscillations
but did not explore their relationship to the collapse.

\citet{2000Hudson} proposed that when a flare or coronal mass ejection (CME)
removes energy from a volume of corona,
 the surrounding magnetic field should collapse or ``implode''
\citep[see also][]{2007JanseLow,2008Hudson}.
This can be understood by noting the equivalence between 
magnetic energy density and magnetic pressure:
thus, a reduction in magnetic energy density creates a 
partial magnetic vacuum that draws in surrounding plasma and magnetic field.
Documented signatures of coronal collapse include contracting magnetic loops
\citep{2006Khan,2009Liu,2009LiuWang,2010LiuWang,2012Liu,Paper1,2014Shen} 
while associated pertubations of the photosphere may be visible as
rapid irreversible changes to the photospheric magnetic field that 
lag the start of the flare by a few minutes
\citep{2005SudolHarvey,2010WangLiu,2010PetrieSudol,2012Johnstone}. 

Separately, transverse oscillations of coronal loops have received much attention 
since their first direct detection \citep{1999Aschwanden,1999Nakariakov}.
The Atmospheric Imaging Assembly (AIA)
on board NASA's Solar Dynamics Observatory (SDO) satellite \citep{2012AIA,2012SDO} 
continues the stream of reported oscillation events 
and further stimulates interest in these phenomena
\citep[e.g.][]{2011AschwandenSchrijver,2012White,2012WhiteVertical,
2012WangT,2012Gosain,2013Verwichte}.
Oscillations of this type are normally interpreted as fast kink magnetohydrodynamic (MHD) 
modes of cylinder-like structures \citep{1983EdwinRoberts}
and, on this basis, procedures have been developed 
to deduce coronal parameters from wave properties
\citep[e.g.][]{1999Nakariakov,2001NakariakovOfman}. 
Reviews on MHD waves and coronal seismology are given by
\citet{2005NakariakovVerwichte,2005DeMoortel,2007Banerjee,2012DeMoortelNakariakov}.

Oscillations and implosions have been noted together in two events
other than the one considered here.
Using SDO, \citet{2012Sun} and \citet{2012Gosain} 
noted loop oscillations during and after a coronal collapse associated with the X2.2 flare SOL2011-02-15T01:44,
with \citet{2012Gosain} estimating oscillation periods and applying seismological formulas 
to estimate coronal Alfv\'en speeds and magnetic field strengths.
Their main focus, however, was the collapse itself
and no conjectures were made about the relationship 
between the collapse and the oscillations.
A causal connection has been suggested by \citet{2010LiuWang},
who observed active region NOAA 10808 with TRACE \citep{1999Trace} 
at the time of SOL2005-09-08T21:05 (X5.4).
They suggested that oscillations could be the result of a contracting loop interacting 
(colliding) with underlying loops, and stated that 
implosions should therefore be considered as an exciter of transverse loop oscillations.
We shall propose a different idea as to the origin of the oscillations, 
which is motivated by the higher quality time series available for our event.

This paper is organised as follows. 
Section \ref{sec:data} presents the observation that motivates this paper.
In Sect.~\ref{sec:excitation}, we propose that loop oscillation and contraction in this event are different aspects of a single
response to the removal of magnetic energy from the corona.
Three types of response are identified and this explains why some loops oscillate while others do not.
The paper concludes with a discussion in Sect.~\ref{sec:disc} \ed{and a summary in Sect.~\ref{sec:conc}}.


\section{Motivating event}\label{sec:data}
On 9 March 2012, the M6.4 class flare SOL2012-03-09T03:53
occurred in active region NOAA 11429, with flare onset recorded at 03:36:45 UT.
The flare was noteworthy for the accompanying coronal collapse, 
which was very clearly captured by SDO/AIA (Paper~I).
It was also associated with coronal dimming in the active region, 
the departure of a coronal mass ejection and a global EUV wave.

Two types of oscillations were excited by the flare.
Hot plasma close to the flare core appeared to undergo 
compressive oscillations with a timescale of approximately 60 seconds.
These manifested as quasi-periodic pulsations in the EUV, SXR and HXR time-series
and may have been standing slow waves
(see Paper~I for the observations and more detailed discussion).
In addition, images collected by SDO/AIA at 171 \r{A}
reveal transverse (kink) oscillations with periods between two and five minutes
in the motions of \ed{various} coronal loops overlying the flare core.
These oscillations began when the loops started to contract 
and continued after the contraction.

\begin{figure}
 \centering
 \includegraphics[width=8.8cm]{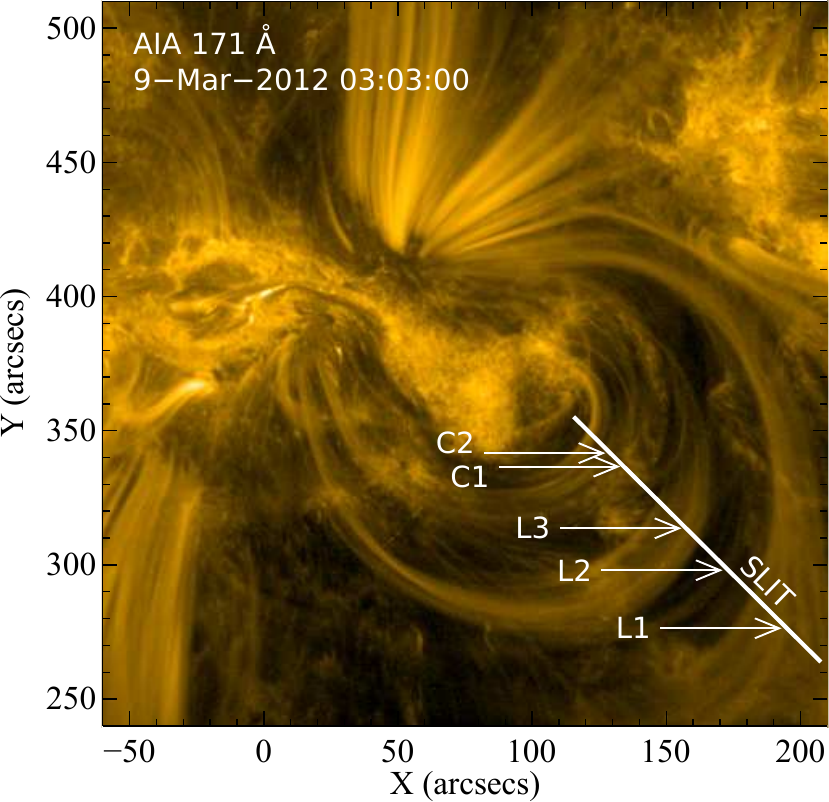}
 \caption{SDO/AIA 171 \r{A} image of active region NOAA 11429 at 03:03:00 UT on 9 March 2012.
          Labelled arrows indicated the loops studied in this paper and the solid line
          represents the artificial slit used to analyse their motions.
          Intensity is displayed using the standard AIA 171 \r{A} colour table.}
 \label{fig:overview}
\end{figure}

\begin{figure}
 \centering
 \includegraphics[width=8.8cm]{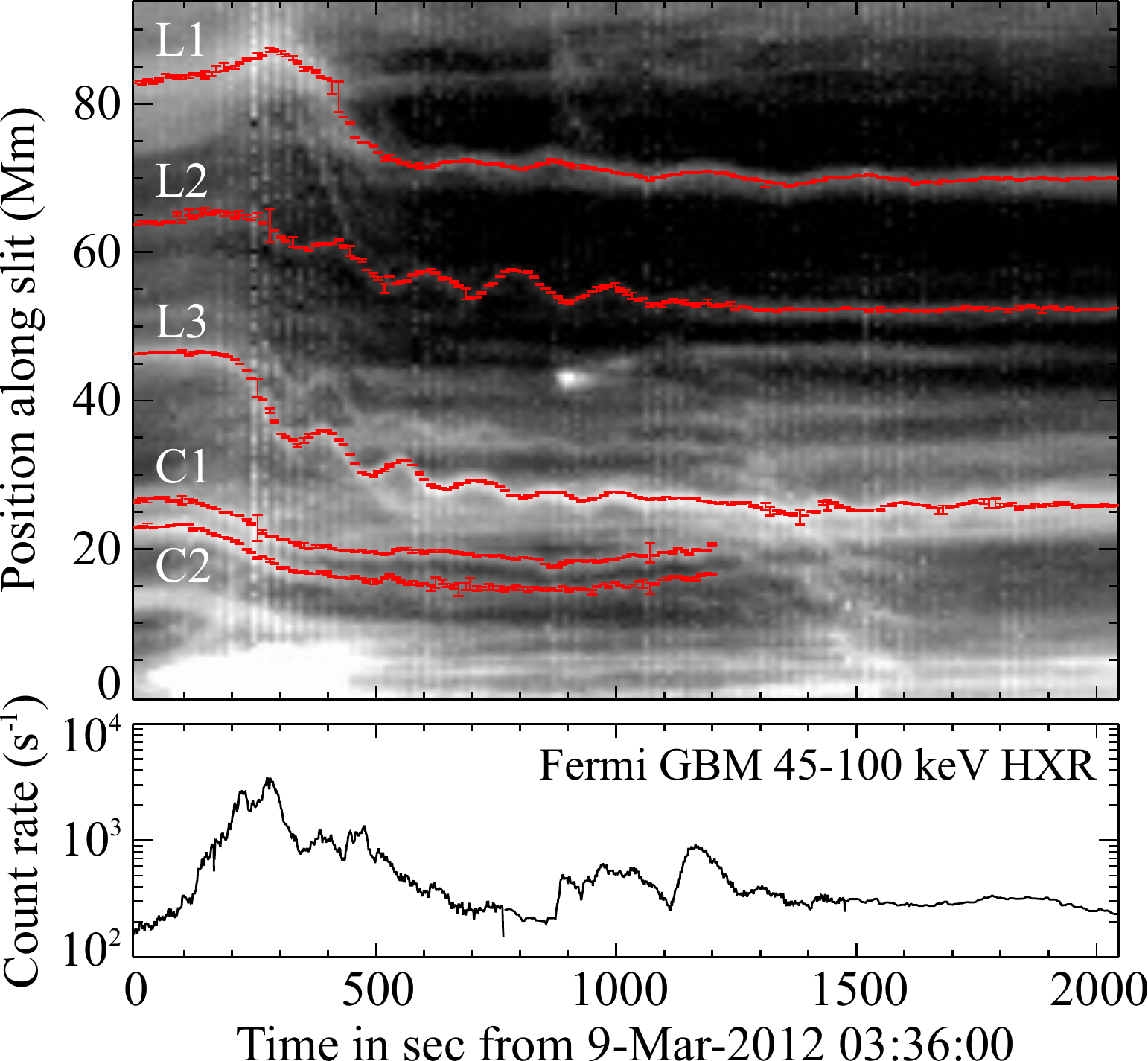}
 \caption{(top) Time-distance image showing contraction and 
          oscillation of coronal loops.
          Intensity has been scaled logarithmically to highlight the loops and
          fitted positions are overplotted with  $1\sigma$ error bars.
          (bottom) Fermi GBM count rate in the HXR 45-100~keV band.
          }
 \label{fig:compare}
\end{figure}

The present paper is motivated by the motions of coronal loops overlying the flare arcade,
all of which contract during the flare, and some of which oscillate as well.
Figure \ref{fig:overview} indicates \ed{five representative} loops of interest.
To analyse their motion, the artificial slit shown in Fig.~\ref{fig:overview}
was used to construct a time-distance image of intensity, 
which is presented in Fig.~\ref{fig:compare} together with fitted loop positions.

The fitted loop positions were obtained by manually identifying spatial ranges
where intensity along the slit is due to a single loop plus some slowly varying background,
then fitting a Gaussian plus a straight line using the Levenburg-Marquardt least-squares method \citep{2009Markwardt}.
The centre of each Gaussian was taken as the loop position
and the $1\sigma$ position uncertainty was estimated using the covariance matrix,
the chi-squared measure of goodness of fit and the number of points fitted.
The typical $1\sigma$ uncertainty is 20\% of the pixel size.

The shortest, lowest loops (C1 and C2) both contract without detectable oscillation,
while the three longer, upper loops (L1, L2 and L3) oscillate significantly 
during and after the contraction.
Their periods are 290~s (L1), 190~s (L2) and 150~s (L3).

The start and duration of the coronal collapse 
coincide with the start and duration of the flare impulsive phase,
subject to time delays that correspond to outward propagation of a signal at 300~km/s
(see Paper~I for the timing analysis).
This relationship can be seen in Fig.~\ref{fig:compare} by comparing the loop motions
with the bright EUV emission at the base of the slit and with the 
45-100~keV HXR count rate from the Fermi 
Gamma-ray Burst Monitor \citep[GBM,][]{2009FermiGBM} 
(bottom panel of Fig.~\ref{fig:compare}). 
L1 has a slow rise phase before collapsing, which may be connected to a departing CME,
however, this slow rise is significantly less apparent for L2 and not discernible in L3.
\ed{A fourth loop that oscillates and contracts is evident just above L3.
The motion of this fourth loop is very similar to that of nearby L3, which is brighter and easier to track.
We therefore use L3 as the representative of motions in its part of the active region,
which is sufficient for development of the discussion in this paper.}
EUV emission in the active region makes it difficult 
to track C1 and C2 reliably beyond 1200~s, 
but that is not a problem since their contraction is completed before then.

The questions we wish to answer are what connects contraction and transverse oscillation for the upper loops,
and why do some loops oscillate but not others?


\section{Combined model of contraction and oscillation}\label{sec:excitation}

\begin{figure}
 \centering
 \includegraphics[width=8.8cm]{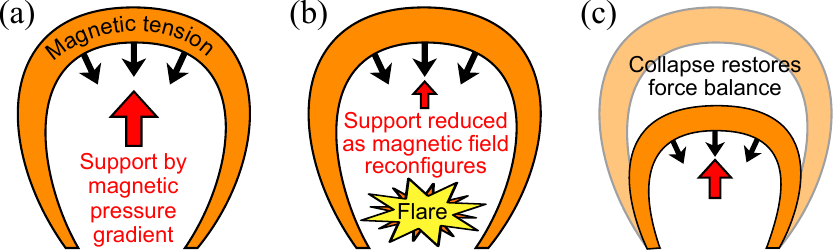}
 \caption{Cartoon of the ``removal-of-support'' mechanism.
          \ed{(a) In the initial equilibrium, inward magnetic tension
          is balanced by a supporting outward force from the magnetic pressure gradient.}
          (b) A flare below the loop removes magnetic energy from the corona, 
          thereby reducing the loop's support \ed{while the tension force is initially unchanged}.
          (c) Unbalanced forces accelerate the loop
          and it moves toward a new position where force balance is restored. 
          }
 \label{fig:collapse}
\end{figure}

\begin{figure}
 \centering
 \includegraphics[width=8.8cm]{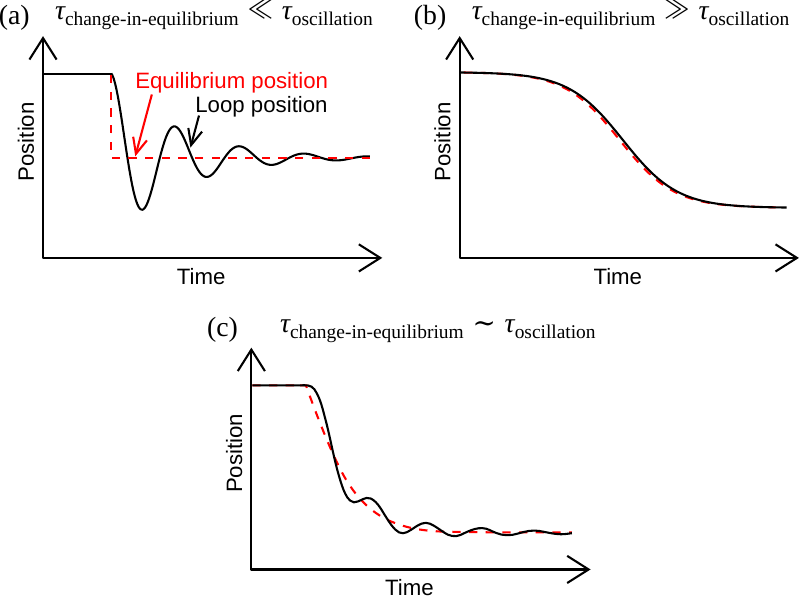}
 \caption{Classes of ``collapse and oscillation'' that result 
           from removing support from loop in different ways.
          The black (solid) line in each plot shows the loop position as a function of time
          and the red (dashed) line shows the equilibrium position, 
          which is moved to drive the system.
          (a) Impulsively excited loop oscillations result if support is removed
           rapidly compared to the oscillation period.
          (b) Gradual displacement is produced if support is removed 
          slowly compared to the oscillation period.
          (c) Oscillation during collapse is seen if support is removed 
          on a time scale broadly similar to the oscillation period.
          }
 \label{fig:excitation}
\end{figure}

We propose that the contraction and oscillation of loops seen in Fig.~\ref{fig:compare} 
are two aspects of a single response to removal of magnetic energy 
from the coronal volume below the loops.

\ed{It is widely thought \citep[e.g.][]{lrsp_benz,lrsp_shibata,2013Su} 
that the energy required to power flares
is released from magnetic fields in the corona by rapid magnetic reconfiguration,
which may be enabled by magnetic reconnection.
The direct coronal magnetic field measurements required to conclusively prove this
are not presently possible, however, less direct estimates of changes to the coronal free magnetic energy 
using magnetic extrapolations generally support this view \citep{2008Schrijver,2010Jing,2012Sun,2014Malanushenko,2014Liu}.
The hypothesised rapid reconfiguration of the coronal magnetic field implies quick changes to the 
$\mathbf{j}\times\mathbf{B}$ Lorentz force acting on the coronal plasma,
and these changes should have a variety of consequences,
including coronal collapse and changes to photospheric magnetic fields,
both of which are observed (references in Sect.~\ref{sec:intro}),
as well as, we argue below, leading to transverse oscillations of contracting loops.}

Figure \ref{fig:collapse} sketches the sequence producing a coronal collapse.
Prior to the flare, the loops are in a stable configuration.
In this initial equilibrium, the magnetic tension \ed{$(\mathbf{B}\cdot\nabla)\mathbf{B}/\mu_0$} component of the 
Lorentz force pulls magnetic loops \ed{inward}
but this is opposed by \ed{outward} forces that support the loop (Fig.~\ref{fig:collapse}a).
In the low-$\beta$ corona, equilibria are well-approximated by magnetic force balance,
so the dominant force supporting loops against contraction
is reasonably considered to be \ed{the outward} magnetic pressure gradient \ed{$-\nabla(B^2/2\mu_0)$
associated with the decrease of magnetic field strength with increasing distance from the core of the active region}.
When a flare occurs, energy for the flare is removed from the coronal magnetic field
and converted to other forms e.g. flare radiative emissions.
The \ed{local} decrease in magnetic energy corresponds to a decrease 
in the magnetic pressure of the affected volume.
\ed{At this point, if we assume that a nearby loop, not involved in reconnection,
has not yet moved and that the magnetic field inside it is unchanged,
then the magnetic tension force for this loop is unaffected by the flare.
However, the magnetic pressure gradient in the loop's vicinity has changed.
Consequently, the forces acting on the loop become unbalanced (Fig.~\ref{fig:collapse}b)
and the loop must move towards the energy release to restore force balance (Fig.~\ref{fig:collapse}c).}
These arguments are in important respects 
equivalent to those advanced by \citet{2000Hudson}.
Similar events will also occur if a CME expels magnetic field from the active region,
e.g. \citet{2014Shen}.

\ed{Since coronal loops have inertia, 
they do not respond instantaneously to changes in their environment.
Instead, they behave like oscillators.
This can be justified physically by noting that 
magnetic restoring forces act to return a displaced coronal loop to the shape/position in which it is in equilibrium.
These forces depend on the instantaneous shape of the loop and on its environment but not its velocity,
and generally increase in magnitude with increasing loop displacements.
If the force perturbations acted in the opposite sense, 
i.e. to accelerate the loop away from the equilibrium position, 
then the loop would be unstable, which is clearly not the case 
for the loops identified in Fig.~\ref{fig:overview} since they were observed over several hours
and survived perturbation by the flare.
When action of restoring forces is combined with inertia, 
standing waves are possible, 
which make the loop displacement at any location oscillate in time.
A more detailed justification for treating coronal loops as oscillators, 
based on the MHD equations, is given in Appendix \ref{app:osc_eq}.}

\ed{For damped oscillations about a fixed equilibrium position, a simple harmonic oscillator obeys}
\begin{align}
 \frac{\mathrm{d}^2x}{\mathrm{d}t^2}+\omega^2(x-x_0)+2\omega\kappa\frac{\mathrm{d}x}{\mathrm{d}t}&=0,\label{eq:toy_minus}
\end{align}
\ed{where $x$ represents displacement of the oscillator, 
$x_0$ is a constant specifying the equilibrium position,
$t$ is time, $\omega$ is the frequency of the corresponding undamped oscillator
 and $\kappa$ is the damping ratio
($\kappa<1$ for underdamping, $\kappa=1$ for critical damping 
and $\kappa>1$ for overdamping).
The same equation can also be rigorously derived for standing kink oscillations 
of particular equilibria as outlined in Appendix \ref{app:osc_eq}.
Here, the damping term is included to represent decay mechanisms such as resonant absorption 
\citep{2002RudermanRoberts,2002Goossens}:
it does not affect the conclusions of this paper, but we include it because
the observed oscillations (Fig.~\ref{fig:compare}) clearly do decay.
We also note that the frequency of oscillation for curved coronal loops
depends (weakly) on the polarisation of the oscillation with respect to the loop's curvature
\citep{2004VanDoorselaere,2004Wang}, 
so $\omega$ should be set according to the polarisation.}

\ed{The purpose of this paper is to show that time-dependent changes 
to the equilibrium position of an oscillator lead to the precisely the 
observed oscillation and displacement behaviours
seen in Fig.~\ref{fig:compare}.
We therefore modify Eq.~(\ref{eq:toy_minus}) 
by making $x_0$ a function of time, $x_0(t)$.
This modification is introduced in an \textit{ad hoc} manner 
rather than by derivation from the full MHD equations,
but it will be justified \textit{a posteriori} by comparison of the 
resulting solutions to the observed loop motions.
This means that we will use the equation}
\begin{align}
 \frac{\mathrm{d}^2x}{\mathrm{d}t^2}+\omega^2(x-x_0(t))
 +2\omega\kappa\frac{\mathrm{d}x}{\mathrm{d}t}&=0,\label{eq:toy}
\end{align}
\ed{as a guide to the expected coronal loop motions,
emphasising that real loops and this model equation 
share the key physical properties of inertia, 
oscillation and changing equilibrium.}

\ed{Three different types of response are found 
depending on how the loop's period of oscillation 
compares to the time scale over which support is removed.
These are shown in Fig.~\ref{fig:excitation}.
Driving functions and parameters that produce examples of each behaviour
can be found in Appendix \ref{app:sketch_driving}.}

Figure \ref{fig:excitation}a illustrates the type of motion that results if support is removed 
rapidly compared to the period of oscillation.
The solid black line shows the loop position and
the dashed red line shows the equilibrium position.
At some time, the equilibrium steps to a new value 
and the loop accelerates away from its original location.
On reaching the new equilibrium position, the loop overshoots, 
and thereafter oscillates about the new equilibrium.
After the initial excitation, the amplitude decays
due to physical damping or other decay processes such as resonant absorption.
In the \ed{model equation}, making $x_0$ a step function leads to a 
straightforward analytic solution
where $x$ is a cosine multiplied by an exponentially decaying envelope.
This scenario can be labelled as an impulsively excited loop oscillation.

The opposite extreme is illustrated in Fig.~\ref{fig:excitation}b.
Here, the equilibrium changes on a time scale 
that is much longer than the period of oscillation.
In this case, the loop's position is always close to the equilibrium position,
although lagging it slightly.
The outcome is that the loop effectively passes through a series of equilibria
and no oscillation would be detected.
This can be referred to as gradual displacement.

The final type of response applies to the intermediate case
in which the equilibrium position changes on a time scale 
broadly similar to the period of oscillation
(i.e. not differing by more than an order of magnitude).
An example of this is shown in Fig.~\ref{fig:excitation}c.
When the equilibrium position starts to move, the loop, having a certain inertia,
requires time to accelerate before it can follow the equilibrium position,
which changes as rapidly as dictated by removal of magnetic energy/pressure.
The loop continues to accelerate until it overtakes the equilibrium position
and the direction of the acceleration is reversed.
Thus, the loop oscillates as in the impulsive case (Fig.~\ref{fig:excitation}a)
but this time the oscillation is superimposed 
with the collapse/displacement that excited it.

Comparing the observations shown in Fig.~\ref{fig:compare}
to the responses illustrated in Fig.~\ref{fig:excitation}, the lower lying loops (C1 and C2) 
fit the pattern of ``gradual displacement''  sketched in Fig.~\ref{fig:excitation}b.
The motion of L3 could serve as an archetype of ``oscillation during collapse'' 
scenario of Fig.~\ref{fig:excitation}c and L2 also belongs to this category. 
Finally, L1 appears to be at the borderline between ``oscillation during collapse''
and the ``impulsively excited loop oscillation'' scenario of Fig.~\ref{fig:excitation}a.
A more distinct example of ``impulsively excited loop oscillation'' 
associated with coronal collapse may be found in the uppermost loop
studied by \citet{2012Gosain} for SOL2011-02-15.
The ordering of the different types of motion also agrees with the model since
loop period increases with loop length from C2 to L1.

Experiments with the \ed{model equation} have shown the greatest amplitude oscillations 
are obtained when the change in equilibrium position is initially sharp,
thereby allowing the equilibrium position to rapidly pull away from the loop
and subjecting the loop to a greater acceleration than if the 
equilibrium position changed more smoothly.
Further detail is given in Appendix \ref{app:sharpness}.
This indicates that oscillations are more likely \ed{to be detected}
if the process removing magnetic energy from the 
corona (presumably magnetic reconnection) switches on over 
a time scale much shorter than the period of oscillation, 
as oppose to ramping up over several oscillation periods.


\section{Discussion}\label{sec:disc}

Loop motions of the type described here are interesting as some of the most
direct evidence for rapid reduction of coronal magnetic energy during flares.
They also offer some interesting diagnostics.

First of all, \ed{our model gives} initial and long term displacements toward 
the coronal volume from which magnetic energy has been removed.
The loops studied in this paper move toward the underlying flare arcade,
suggesting that the energy fuelling the flare had been stored in 
the low corona close to the arcade \ed{\citep[also see]{2012Sun}.
This is reasonable since the core of active region is where the 
coronal magnetic fields are strongest and therefore can most easily
build up significant free magnetic energy.
It is also consistent with the observation that changes to the non-radial photospheric magnetic field  at the time of flares,
which indicate a change in coronal free magnetic energy,
are usually concentrated near the polarity inversion line,
also suggesting that flares are typically powered by free magnetic energy stored in the low corona.}

Our model allows for loop motions other than vertical:
in particular, if a flare were to remove energy from 
a volume at the same height as the non-flaring magnetic loops, then
the loops would move perpendicular to their plane as
force imbalance accelerates them towards the flare volume.
That would excite oscillations perpendicular to each loop's plane rather than within it
\ed{(see \citet{2004Wang} for discussion of differences between 
in-plane and out-of-plane kink oscillations).}
\citet{2013White} recently presented an interesting event that fits this pattern.
Following an M1.4 class flare, two coronal loops were seen to oscillate, 
initially moving toward one another.
A net inward displacement of the loops also occurred during their oscillation.
This behaviour is consistent with our model if magnetic energy 
was removed from the volume between these loops.
Indeed, post-flare loops for this event do form between the eastern footpoints 
of the much longer oscillating loops, in keeping with such a picture.

As well as indicating volumes where magnetic energy has decreased,
loop motions provide information on how rapidly magnetic energy was 
removed and how rapidly the energy release switched on.
Similar information can already be gained from HXR, EUV and radio emissions,
however loop motions are potentially valuable as an additional diagnostic
since they reflect changes to coronal magnetic energy without assumptions 
about conversion to other forms.
For the event considered in this paper, inspection of Fig.~\ref{fig:compare} shows that
most of the collapse occurred within about 300~s, 
with the entire contraction completed within 900~s.
The logical inference is that the removal of magnetic energy also followed this pattern.
These values are consistent with what one would deduce 
from the Fermi GBM {45-100~keV} HXR emission (bottom panel of Fig.~\ref{fig:compare}).
Meanwhile, the large amplitude of the oscillations suggests that the switch-on time 
for the energy release is at most a quarter of the shortest observed period, i.e. 40~sec
(see Appendix \ref{app:sharpness} for analysis).
This time scale is comparable to the shorter rise times of the HXR emission.

The findings of this paper are also of interest to the study of 
coronal oscillations and seismology.
In particular, removal of magnetic energy from near a coronal loop
joins other mechanisms for excitation of kink oscillations,
such as interaction with a fast wave shock (blast) wave \citep{1999Nakariakov,2009Pascoe},
reconnection outflow \citep{2012WhiteVertical},
or vortex-shedding in response to a flow  relative to the loop \citep{2009Nakariakov}.
For the event in this paper, the other mechanisms can be ruled out because 
the loops initially move toward the flare rather than away from it 
(ruling out the blast mechanism),
the loops exist before the flare/eruption and remain cool 
(in contrast to post-reconnection loops),
and the oscillations have their maximum amplitude at the start of the collapse
(indicating that vortex-shedding is not responsible here).
Comparative motion of different loops also indicates a driving signal 
moving away from the flare at around 300~km/s,
which is consistent with information about the removal of 
magnetic energy propagating out from the flare site at the fast speed 
(for low-$\beta$ plasma the fast speed 
is approximately the Alfv\'en speed) but seems inconsistent with driving from above,
e.g. by the departing CME.

In what situations will removal-of-support excite oscillations 
and how common do we therefore expect these events to be?
Firstly, the mechanism requires that magnetic energy is converted 
on a time scale broadly comparable to 
or shorter than the oscillation period of the loop.
Taking the impulsive phase of a flare as a proxy for removal of coronal magnetic energy,
flare impulsive phases typically last between 
several tens of seconds and tens of minutes \citep{2011FletcherSSR}.
Meanwhile, examples of transverse loop oscillations collected by 
\citet{2002Aschwanden} and \citet{2012White}
show periods in the range 1.7 to 33 minutes
(the lower limit is probably observational since SDO/AIA 
has a mean cadence of 12 seconds, excellent by historical standards,
and a minimum of six points per period is required to convincingly resolve an oscillation).
In general terms, then, the typical duration of the flare impulsive phase
is comparable to the lower end of loop periods reported in the observational literature,
which suggests coronal collapse should commonly excite oscillations.
A significant proportion of these will be of the ``oscillate during collapse'' type
while the longest period loops will better fit the ``impulsively excited'' template.
A second consideration is the amount of magnetic energy removed from the corona.
This paper has referred to events with flares of class M1.4 \citep{2013White},
M6.4 (Sect.~\ref{sec:data}), X2.2 \citep{2012Sun,2012Gosain} and X5.4 \citep{2010LiuWang},
so a broad range of M and X flares seem to excite oscillations in this way,
the main difference being that larger flares perturb more of their active region 
and produce larger loop displacements. 
Finally, we have demonstrated that large amplitude oscillations are
favoured by rapid switch-on of the energy release.
This may be the limiting factor, hence we recommend 
that anyone seeking new examples first 
look to flares where the impulsive phase commenced suddenly.

Since many loops in the same active region may be excited by coronal collapse
(as in the SOL2012-03-09 and SOL2011-02-15 events),
oscillations excited by coronal collapse could be particularly useful 
for multi-loop coronal seismology,
e.g. using oscillations of several loops to build a picture of the 
coronal Alfv\'en speed throughout an active region.
This application is \ed{complicated} slightly by the need 
to separate the loop motion into oscillation and displacement components, however, 
we have found this issue surmountable, as will be shown in future work.

\ed{Finally, the conclusions of this paper rely only on coronal loops having inertia, 
being natural oscillators 
and having equilibrium shapes/positions that change during flares.
These points are well established observationally and no further assumptions are required, 
so we expect the simple analysis in this paper to be a good guide to the qualitative behaviour.
Nonetheless, it could still be informative to perform 3D MHD simulations 
of the loss-of-support mechanism,
which would provide further validation and add details about the precise manner 
in which the loop's equilibrium shape/position changes in response to 
the removal of coronal magnetic energy.
We anticipate such simulations in the near future.}


\section{\ed{Conclusions}}\label{sec:conc}

This paper has argued that coronal loop contraction and oscillation in flares 
can occur as part of a single response to the removal of magnetic energy from the corona.
The corresponding framework successfully connects oscillation and contraction,
and explains why some loops oscillate while others do not.

The key ideas are:
\begin{enumerate}
\item Conversion of magnetic energy reduces magnetic pressure, 
thereby changing the position in which nearby coronal loops are in equilibrium.
\item Since coronal loops possess mass, 
they cannot respond instantly to changes in their environment.
\item Comparison of a loop's natural period of oscillation to the time scale of the 
energy release determines the type of motion that will result.
Loops with periods much longer than the time scale of the energy release
oscillate around their new equilibrium position with 
an initial amplitude set by the displacement between the old and new equilibrium positions;
loops with periods much shorter than the energy release time scale 
contract with negligible oscillation;
and loops with periods broadly comparable to the energy release time scale 
exhibit both behaviours at once.
\end{enumerate}

In conclusion, contraction or displacement of coronal loops and transverse loop oscillations
are closely related in solar flares, and these motions reveal 
much about the corona and its dynamics,
especially rapid decreases in coronal magnetic energy.

\begin{acknowledgements}
   We gratefully acknowledge support from the Royal Commission for the
   Exhibition of 1851 through their Research Fellowship scheme (AJBR),
   the European Commission through HESPE 
   (FP7-SPACE-2010-263086) (PJAS, LF)
   and the Science and Technology Facilities Council  through grants 
   ST/I001808 (PJAS, LF) and ST/L000741/1 (LF) to the University of Glasgow,
   as well as ST/K000993/1 (AJBR) to the University of Dundee.
   AIA data were supplied courtesy of SDO (NASA) and the AIA science team. 
\end{acknowledgements}

\bibliographystyle{aa}
\bibliography{unified}

%

\appendix
\section{\ed{Coronal loops as harmonic oscillators}}\label{app:osc_eq}
\ed{The treatment of coronal loops as harmonic oscillators can be rigorously justified for particular equilibria.
For example, neglecting gravity, gas pressure, resistivity and viscosity for simplicity
and assuming linear perturbations about a static potential equilibrium, 
the momentum equation can be written as}
\begin{eqnarray}
  \frac{\partial^2 \mathbf{\xi}}{\partial t^2} &=& \frac{1}{\mu_0\rho_0}\left(\nabla\times\mathbf{b}\right)\times\mathbf{B}_0
\end{eqnarray}
\ed{where $\mathbf{\xi}$ is the plasma displacement vector,
$\mathbf{b}$ is the magnetic field perturbation,
$\rho_0$ is the equilibrium density and
$\mathbf{B}_0$ is the equilibrium magnetic field which we have assumed curl-free.
Similarly, the time-integrated induction equation is
\begin{eqnarray}
  \mathbf{b} &=& \nabla\times\left(\mathbf{\xi}\times\mathbf{B}_0\right).
\end{eqnarray}
We consider a straight equilibrium magnetic field by setting $\mathbf{B}_0=B_0\mathbf{\hat{z}}$ 
($B_0$ is constant to give magnetic pressure balance in the equilibrium).
Then, performing some algebra,
the governing equations can be combined to give
\begin{eqnarray}\label{eq:osc:wave}
  \frac{\partial^2 \mathbf{\xi}_\perp}{\partial t^2} &=& 
    v_A^2\left( \frac{\partial^2 \mathbf{\xi}_\perp}{\partial z^2}+ \nabla_\perp\left(\nabla\cdot\mathbf{\xi}_\perp\right)\right)
    \equiv\mathcal{W}\mathbf{\xi}_\perp,
\end{eqnarray}
where $v_A=B_0/\sqrt{\mu_0\rho_0}$ is the Alfv\'en speed
and $\mathcal{W}$ is a spatial differential operator defined by the equation above.
In a medium with uniform Alfv\'en speed, 
incompressible solutions (with $\nabla\cdot\mathbf{\xi}_\perp=0$) 
to Eq.~(\ref{eq:osc:wave})  are Alfv\'en waves,
while the divergence of Eq.~(\ref{eq:osc:wave}) gives the governing equation 
for fast waves in cold plasma.
Equation (\ref{eq:osc:wave}) is also valid when the Alfv\'en speed is a function of position,
as happens in equilibria supporting kink waves.}

\ed{Standing wave solutions are obtained when $\mathbf{\xi}_\perp$ 
has separable time and spatial dependences, i.e.
\begin{eqnarray}
  \mathbf{\xi}_\perp &=& T(t)\mathbf{X}(\mathbf{x}).
\end{eqnarray}
Substituting this form into Eq.~(\ref{eq:osc:wave}), we find
\begin{eqnarray}
  \frac{1}{T(t)} \frac{ \mathrm{d}^2T(t) }{ \mathrm{d}t^2 }&=&
  \frac{\mathcal{W}X_1(\mathbf{x})}{X_1(\mathbf{x})}=
  \frac{\mathcal{W}X_2(\mathbf{x})}{X_2(\mathbf{x})}=-\omega^2
\end{eqnarray}
where $X_1$ and $X_2$ are the components of $\mathbf{X}$ in the chosen coordinate system
and we have introduced $-\omega^2$ as the separation constant.
It immediately follows that 
\begin{eqnarray}
  \frac{\mathrm{d}^2 T}{\mathrm{d} t^2} + \omega^2 T &=& 0,\label{eq:osc:osc}
\end{eqnarray}
which is the ordinary differential equation for a harmonic oscillator,
while
\begin{eqnarray}
  \mathcal{W}X_j(\mathbf{x})&=&-\omega^2X_j(\mathbf{x}), \quad j\in\{1,2\}\label{eq:osc:eigen}
\end{eqnarray}
implies that $X_1$ and $X_2$ are eigenfunctions of $\mathcal{W}$, each having eigenvalue $-\omega^2$.
Oscillations appear for real $\omega$, 
in which case Eq.~(\ref{eq:osc:osc}) can be scaled so that $T$ represents, 
e.g., the displacement of the loop apex from its equilibrium position.
If $\omega$ is imaginary then the equilibrium can be unstable, 
but we exclude that possibility here for consistency with our observations.
The final possibility for $\omega^2$ is that it could potentially be complex, 
which may introduce decay alongside oscillation as discussed later.}

\ed{The existence of suitable eigenfunctions of $\mathcal{W}$ is expected on the basis 
that standing transverse oscillations of coronal loops are observed, 
including in the event presented in Sect.~\ref{sec:data}.
We also provide mathematical justification by considering kink modes for 
the well-known magnetic cylinder model of \citet{1983EdwinRoberts}.
In this model, the equilibrium Alfv\'en speed is specified in cylindrical coordinates $(r,\theta,z)$ as
\begin{eqnarray}
  v_A(r) &=& \begin{cases} 
    v_{A0}, &r<a \\ 
    v_{Ae}, &r>a \end{cases},
\end{eqnarray}
with $v_{Ae}>v_{A0}$ so that the inner region of radius $a$ acts as a waveguide for fast waves, 
giving rise to body modes.
Since we have $B_0$ constant for equilibrium magnetic-pressure balance,
the Alfv\'en speed profile corresponds to
a density enhancement inside the cylinder.
We also assume that there are line-tied boundaries at $z=0$ and $z=L$,
which gives rise to standing waves with wave number $k_z$,
the fundamental harmonic having $k_z=\pi/L$.
The spatial eigenvalue equations (\ref{eq:osc:eigen}) can be solved for this equilibrium 
using the approach described by \citet{1983EdwinRoberts},
which yields $X_r(r,\theta,z)$ and $X_\theta(r,\theta,z)$ with radial dependences in terms of Bessel functions.
Meanwhile, the eigenvalue is determined by 
\begin{eqnarray}
  \omega^2&=&c_p^2k_z^2\label{eq:osc:disp}
\end{eqnarray}
where $c_p(ak_z)$ is the phase-speed of the kink wave.
In the thin-tube limit, $k_z a\ll 1$, the waves become non-dispersive and $c_p$ reduces to the kink speed,
\begin{eqnarray}
   c_k &=& \left(\frac{\rho_0 v_{A0}^2+\rho_e v_{Ae}^2}{\rho_0+\rho_e}\right)^{1/2}.
\end{eqnarray}
Since spatial eigenfunctions have been found and $\omega^2$ is real and positive, 
standing kink waves in the classic magnetic cylinder model
are consistent with a harmonic oscillator view of coronal loops.}

\ed{More generally, standing kink oscillations can decay over time 
due to physical damping by viscosity, resistivity etc.
or due to transfer of energy from the kink mode by resonant absorption 
\citep{2002RudermanRoberts,2002Goossens}.
In these situations, the above approach of separation of variables can still be applied,
leading to an oscillation equation of the form of Eq.~(\ref{eq:osc:osc}).
However, solution of the spatial operator equations now yields a complex frequency, 
$\omega=\omega_r+i\omega_i$.
When this occurs it can be useful to recast the oscillation equation as 
a damped harmonic oscillator with real coefficients:
\begin{eqnarray}
  \frac{\mathrm{d}^2 T}{\mathrm{d} t^2} + \omega_0^2 T
  + 2\omega_0 \kappa \frac{\mathrm{d}T}{\mathrm{d}t}&=& 0,
  \label{eq:osc:damped}
\end{eqnarray}
where $\omega_0$ is the frequency of corresponding undamped oscillator 
and $\kappa$ is the damping ratio.
To see the correspondence, observe that harmonic solutions 
of this new equation with an $\mathrm{e}^{i\omega t}$ time dependence
also have complex frequency, with $\omega_r=\omega_0\sqrt{1-\kappa^2}$ 
and $\omega_i=\omega_0\kappa$
(we will consider underdamped solutions with $0<\kappa<1$
so $\omega_r$ is always real and $\omega_i>0$ giving decay).
This implies that the original oscillation equation 
with complex $\omega$ has the same oscillatory and decay behaviours 
as the damped oscillation equation with real coefficients
set as $\omega_0^2=\omega_r^2+\omega_i^2$ and $\omega_0\kappa=\omega_i$,
where $\omega_r$ and $\omega_i$ are obtained from the spatial eigenvalue equations.}
 
\ed{We conclude by noting that the approach described above is fairly general
and it can be applied to analyses that make fewer simplifying assumptions.
For example, gas pressure was in fact included by \citet{1983EdwinRoberts},
while magnetic curvature, density stratification, flux tube expansion and 
non-circular cross-sections have all been considered since then
\citep[see][]{2003Ruderman,2004VanDoorselaere,2005Andries,2008Ruderman,2015Ruderman}.
In each case, suitable spatial eigenfunctions were obtained 
and a dispersion relation giving $\omega$ was derived,
which determines the real coefficients in the corresponding damped
oscillation equation.}

\section{Solution of the \ed{model equation}}\label{app:sketch_driving}
Our \ed{model equation} produces all three types 
of solution discussed in Sect.~\ref{sec:excitation}.
Writing Eq.~(\ref{eq:toy}) in dimensionless form,
\begin{align}\label{eq:toy_norm}
   \frac{d^2\tilde{x}}{d\tilde{t}^2}+
  4\pi^2(\tilde{x}-\tilde{x}_0(t))+
  4\pi\kappa\frac{d\tilde{x}}{d\tilde{t}}&=0,
\end{align}
where $\tilde{t}=t/\tau$ with $\tau=2\pi/\omega$,
and $\tilde{x}=x/D$ with $D$ an appropriate length scale such as the final displacement.
This equation is equivalent to a pair
of coupled first-order ordinary differential equations,
\begin{align}\label{eq:sys}
  \begin{split}
  \frac{d\tilde{x}}{d\tilde{t}} &= \tilde{v}, \\
  \frac{d\tilde{v}}{d\tilde{t}} &= 
  -4\pi^2(\tilde{x}-\tilde{x}_0(t)) -  4\pi\kappa\tilde{v},
 \end{split}
\end{align}
which are easily integrated using fourth-order Runge-Kutta.
Setting the damping parameter as $\kappa=0.1$ produces decay
similar to the motivating example. 
Solutions of the three types shown in Fig.~\ref{fig:excitation}
can then be obtained using different driving functions for $\tilde{x}_0(\tilde{t})$.
Impulsively excited oscillations (Fig.~\ref{fig:excitation}a)  
result if the system is driven using a step function, e.g.
\begin{align}\label{eq:driving_a}
  \tilde{x}_0(\tilde{t}) &= \begin{cases}
   2, & \tilde{t} < \tilde{t}_c \\
   1, & \tilde{t} > \tilde{t}_c
  \end{cases}  
\end{align}
where $\tilde{t}_c$ is the time of the collapse. 
Examples of gradual displacement (Fig.~\ref{fig:excitation}b)
can be produced using
\begin{align}\label{eq:driving_b}
  \tilde{x}_0(\tilde{t}) &= 
  \frac{1}{2}\left[3-\tanh\left(\frac{\tilde{t}-\tilde{t}_c}{\Delta}\right)\right] ,
\end{align}
with $\Delta\gtrsim1$.
In fact, Eq.~(\ref{eq:driving_b}) can produce any of the three types of response,
depending on the value of $\Delta$,
and converges to Eq.~(\ref{eq:driving_a}) for $\Delta\to 0$.
Nonetheless, the clearest examples of oscillation during collapse (Fig.~\ref{fig:excitation}c)
are  produced when the change in equilibrium starts sharply, 
e.g. under
\begin{align}\label{eq:driving_c}
  \tilde{x}_0(\tilde{t}) &= \begin{cases}
    2, & \tilde{t} \leq\tilde{t}_c \\
    2 - \tanh\left(\left(\tilde{t}-\tilde{t}_c\right)/\Delta\right), & \tilde{t} > \tilde{t}_c
  \end{cases}
\end{align}
with $\Delta=1$.

\section{Effect of switch-on time}\label{app:sharpness}
\begin{figure}
 \centering
 \includegraphics[width=6.6cm]{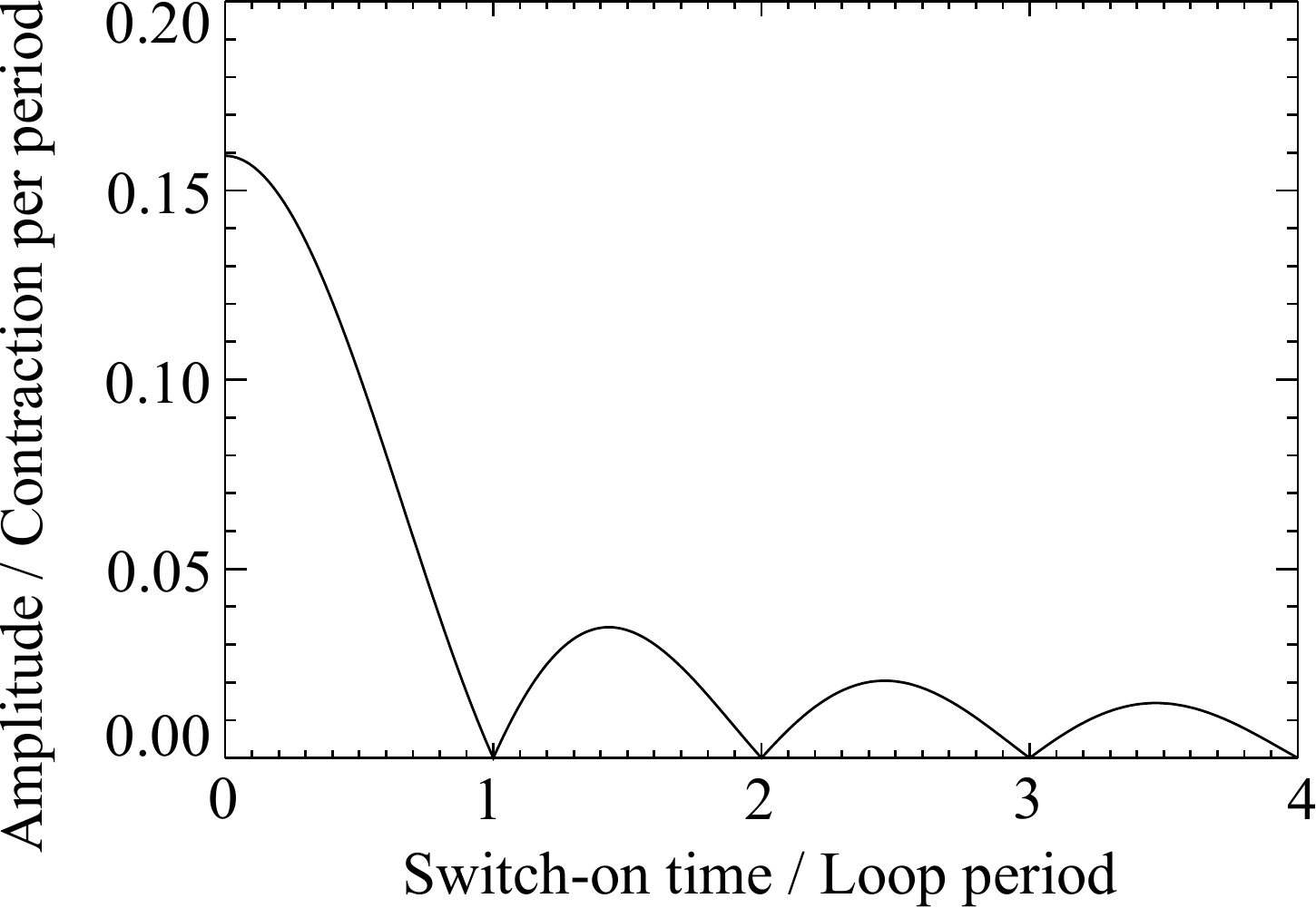}
 \caption{Amplitude of oscillation as a function of the energy-release switch-on time,
          for a test problem in which the equilibrium position accelerates from rest
          to motion at a constant speed.
          The switch-on time (duration of the acceleration) 
          is normalised to the period of oscillation, 
          and the amplitude of oscillation is normalised to the 
          equilibrium displacement per period after the acceleration.
          }
 \label{fig:sharpness}
\end{figure}

It was noted in Sect.~\ref{sec:excitation} that the greatest amplitude 
oscillations are obtained when the change in equilibrium position
is initially sharp.  Physical justification was given in Sect.~\ref{sec:excitation},
but a quantitative demonstration is also provided here.
Using the non-dimensionalised model of Eq.~(\ref{eq:toy_norm}) 
with damping turned off ($\kappa=0$), 
we consider a test problem in which the equilibrium position
accelerates from rest to motion at a constant speed.
Assuming the equilibrium position accelerates at a constant rate
during a switch-on interval equivalent to $\delta$ periods of oscillation,
\begin{align}\label{eq:sharpness_driver}
  \tilde{x}_0(\tilde{t}) &= \begin{cases}
    \tilde{x}_r, & \tilde{t} \leq\tilde{t}_c \\
    \tilde{x}_r-\tilde{t}^2/(2\delta), & \tilde{t}_c < \tilde{t} < \tilde{t}_c+\delta \\
    \tilde{x}_r -\tilde{t}+\delta/2, & \tilde{t}_c+\delta \leq\tilde{t} 
  \end{cases}.
\end{align}
Here, the normalising length has been set to the distance that 
the equilibrium position moves per period in the constant speed phase,
$\tilde{t}>\tilde{t}_c+\delta$.
The specified driver produces oscillations superimposed on a never-ending collapse,
and the amplitude of the oscillations is readily measured from the long term solution.
Figure \ref{fig:sharpness} plots the normalised amplitude for a range of switch-on times,
$\delta$, which is the only free parameter in the dimensionless model.
In this test problem, \ed{dimensional} amplitude is limited only 
by the maximum rate of collapse,
with amplitudes capped at 0.16 times the maximum displacement per period.
Nodes in Fig.~\ref{fig:sharpness} show cases where 
acceleration of the equilibrium position resonates with the loop period
in such a way that long-term oscillations are not produced.
Most significantly, the plot confirms that the largest amplitude oscillations 
are excited when the switch-on time is a small fraction of the oscillation period.

Examining the motion of loop L3 in Fig.~\ref{fig:compare}, 
the inferred equilibrium position shifts by approximately 12~Mm 
during the loop's first period of oscillation.
Multiplying this distance by 0.16 gives an estimate of 2~Mm for the 
maximum amplitude that can be excited by the corresponding rate of contraction.
That estimate is very close to the actual amplitude seen for L3.
Based on these values and the fall-off of amplitude with increasing switch-on time
(Fig.~\ref{fig:sharpness}) we suggest that a quarter of L3's period
(approximately 40~s) is a reasonable upper limit to place on the 
energy-release switch-on time for the SOL2012-03-09 flare.

\end{document}